\documentclass[apj]{emulateapj}
\usepackage{graphicx}

\def\lexp{\mathop{\langle}\nolimits}
\def\rexp{\mathop{\rangle}\nolimits}

\def\B{{\vec B}}
\def\v{{\vec v}}
\def\k{{\vec k}}
\def\khat{{\hat k}}

\def\del{\vec\nabla}

\def\rhobar{{\bar \rho}}

\def\thetaq{{\theta_q}}
\def\deltaq{{\delta_q}}
\def\G{{\mathrm{G}}}
\def\uG{{\mu\G}}
\def\g{\,{\rm g}}
\def\cm{\,{\rm cm}}
\def\km{\,{\rm km}}
\def\s{\,{\rm s}}
\def\Mpc{{\mathrm{Mpc}}}
\def\dD{\delta_{\rm D}}

\def\COSMICS{{\tt COSMICS}}

\begin{document}

\title{Cosmological Structure Formation Creates Large-Scale Magnetic Fields}

\author{E. R. Siegel and J. N. Fry}

\affil{Department of Physics, University of Florida, Gainesville,
FL 32611-8440; \break siegel@phys.ufl.edu, fry@phys.ufl.edu}

\begin{abstract}

This paper examines the generation of seed magnetic fields 
due to the growth of cosmological perturbations.
In the radiation era, different rates of scattering from photons 
induce local differences in the ion and electron density 
and velocity fields.
The currents due to the relative motion of these fluids 
generate magnetic fields on all cosmological scales, 
peaking at a magnitude of $\mathcal{O}(10^{-24} \, \G) $  
at the epoch of recombination.
Magnetic fields generated in this manner provide a promising 
candidate for the seeds of magnetic fields presently 
observed on galactic and extra-galactic scales.

\end{abstract}

\subjectheadings{cosmology : theory --- early universe ---
large-scale structure of universe --- galaxies: magnetic fields}

\section{Introduction}

The presence of magnetic fields on galactic and extragalactic
scales is a major unsolved puzzle in modern astrophysics.
Although the observational evidence supporting the existence of 
magnetic fields in
large-scale structure is overwhelming, there is no consensus as
to their origins.  The standard paradigm for the creation of these
fields is the dynamo mechanism, in which a small initial seed
field is amplified by turbulence (the $\alpha$ effect)
and/or by differential rotation (the $\omega$ effect).
In principle, once a seed field is in place, it should be possible
to follow its evolution and amplification from the collapse of
structure and the operation of any relevant dynamos.

In recent papers, it has been argued that magnetic fields are generated
in the early universe due to induced vorticity 
\citep{Italians:04,GS:05,Japanese:04}.  
In this paper, we put forward a new mechanism for the generation 
of seed fields from cosmological perturbations.
We show that the evolution of cosmological perturbations in the 
pre-recombination era 
produces charge separations and currents on all scales, 
both of which contribute to magnetic fields, in addition to
and independent of any fields generated via vorticity.
These seed fields persist until the onset of significant 
gravitational collapse, 
at which point field amplification and dynamo processes may magnify 
such seeds to the $\mathcal{O}(\uG)$ fields observed today.

The generation of magnetic fields from charge separations and 
currents in the early universe 
is a necessary consequence of structure formation.
This paper calculates the magnitude of these seed fields, and it is shown
that these seed fields may be sufficiently strong to
account for all of the observed magnetic fields in large-scale
structures. The layout of the paper is as follows: 
Section 2 gives an overview of the observational evidence for magnetic
fields along with a brief theoretical picture of their generation.
Section 2 also contains an explanation of the novel idea that the 
early stages of structure formation in a perturbed universe 
generate magnetic fields.
Section 3 presents a detailed treatment of cosmological perturbations,
with a specific view towards the creation and evolution of 
local charge separations and currents.
Section 4 explicitly calculates the seed magnetic fields that 
arise via this mechanism as a function of scale and epoch.
Finally, section 5 compares the results of this mechanism
with competing theories.  It also discusses avenues for future
investigation of this topic, including possible observational
signatures of the fields that would arise from this mechanism.

\section{Magnetic Fields: Background}


Magnetic fields with strength $\sim \uG $ are seen 
in all gravitationally bound or collapsing structures in which the 
appropriate observations have been made \citep{Widrow:02}.
The four major methods used to study astrophysical magnetic fields are
synchrotron radiation, Faraday rotation, Zeeman splitting, and
polarization of starlight.  These observational techniques are
detailed in depth in \citet{Ruz:88}, with Faraday rotation often 
proving the most fruitful of the above methods.

Magnetic fields have been found in many different types of
galaxies, in rich clusters, and in galaxies at high redshifts.
Spiral galaxies, including our own, appear to have relatively
large magnetic fields of $\mathcal{O}(10 \,\uG)$ on
the scale of the galaxy \citep{Fitt:93}, 
with some (such as M82) containing 
fields up to $\simeq 50 \, \uG$ \citep{Klein:88}.  
Elliptical and irregular galaxies possess 
strong evidence for magnetic fields (of order $\sim \uG$) 
as well \citep{Moss:96}, although they are much more 
difficult to observe due to the paucity of free electrons in these 
classes of galaxies.
Coherence scales for magnetic fields in these galaxies, as opposed 
to spirals, are much smaller than the scale of the galaxy, of order
$100$ to $1000$ pc.
Furthermore, galaxies at moderate $(z \simeq 0.4)$ and high 
redshifts $(z \gtrsim 2)$ have 
been observed to require significant $(\sim \uG)$
magnetic fields to explain their observed Faraday rotations
\citep{KPZ:92,Athreya:98}.  Magnetic fields are also observed in 
structures larger than individual galaxies.  The three main types 
of galaxy clusters are those with cooling flows, those with 
radio-halos, and those devoid of both.  Galaxy clusters with cooling 
flows are observed to have fields of $ 0.2 $ {to}  $ 3 \, \uG$ 
\citep{TBG:94}, the Coma cluster (a prime example of a radio-halo
cluster) is observed to have a field strength $\sim 2.5 \, \uG$ 
\citep{Kim:90}, while clusters selected to have
neither cooling flows nor radio halos still exhibit indications of
strong $(0.1$--$1 \,\uG)$ fields \citep{CKB:01}. There even
exists evidence for magnetic fields on
extracluster scales.  An excess of Faraday rotation is observed for 
galaxies lying along the filament between the Coma cluster and the 
cluster Abell 1367, consistent with an intercluster magnetic field 
of $0.2$--$0.6 \, \uG$ \citep{Kim:89}. 
On the largest cosmological scales,
there exist only upper limits on magnetic fields, arising from
observations of the cosmic microwave background 
\citep{BFS:97,Yamazaki:05,Yamazaki:06} and
from nucleosynthesis \citep{CST:94}, setting limits that on scales
$\gtrsim 10 \, \Mpc $, field strengths are $\lesssim 10^{-9} \, \G$.

Observational evidence for magnetic fields is found in galaxies of
all types and in galaxy clusters, both locally and at high redshifts,
wherever the appropriate observations can be made.  A review of
observational results can be found in \citet{Vallee:97}.  The
theoretical picture of the creation of these fields, however, is
incomplete.  Fields of strength $\sim \uG$ can be
explained by the magnification of an initial, small seed field on
galactic (or larger) scales by the dynamo mechanism
\citep{Park:71,Vain:71,Vain:72}.  A protogalaxy (or protocluster)
containing a magnetic field can have its field strength increased
by many orders of magnitude through gravitational collapse
\citep{LC:95,HK:97}, and can also be further amplified via various
dynamos. Dynamos which can amplify a small seed field
into the large fields observed today involve helical turbulence
$(\alpha)$ and/or differential rotation $(\omega)$.  Various
types of these dynamos include the
mean-field dynamo \citep{Steenbeck:66,Moffatt:78,Ruz:88}, the
fluctuation dynamo \citep{Kazantsev:85,Moss:96}, and merger-driven
dynamos \citep{Tribble:93}, among others. However, the dynamo
mechanism does not explain the origin of such seed fields.


While the initial seeds that grow into magnetic fields are
anticipated to be small, they must still come from somewhere
\citep{ZN:83}, as their existence is not explained by the dynamo
mechanism alone.  There are many mechanisms that can produce
small-strength magnetic fields on astrophysically interesting
scales, either through astrophysical or exotic processes
\citep[see][for a detailed review]{Widrow:02}. Exotic
processes generally rely on new physics in the early universe,
such as a first-order QCD phase transition
\citep{Hogan:83,QLS:89}, a first order electroweak phase
transition \citep{BBM:96,SOJ:97}, broken conformal invariance
during inflation \citep{Turner:88,AM:05}, specific inflaton
potentials \citep{Ratra:92}, or the presence of charged scalars
during inflation \citep{CKM:98,Kandus:00,Davis:01}.  Astrophysical
mechanisms, in contrast, are generally better grounded in known
physics, although they have difficulty generating sufficiently
strong fields on sufficiently large scales.  The difference in
mobility between electrons and ions admits, under appropriate
circumstances, the creation of large-scale currents and magnetic
fields.  Seed magnetic fields arising from this difference in
mobility between electrons and ions may be astrophysically generated
via a variety of phenomena, including
radiation-era vorticity \citep{Harrison:70,Harrison:73}, 
vorticity due to gas-dynamics in ionized plasma
\citep{Biermann:50,Pudritz:89,SNC:94,Kulsrud:97,Gnedin:00}, 
stars \citep{Syro:70}, or from active galactic nuclei
\citep{Hoyle:69}.  Although there are many candidates for
producing the seed magnetic fields required by the dynamo
mechanism, none has emerged as a definitive solution to the puzzle
of explaining their origins.



The mechanism proposed in this paper is that seed magnetic
fields are generated by the scattering of photons with charged
particles during the radiation era.  Unlike the mechanism of
\citet{Harrison:70,Harrison:73}, which is disfavored
\citep{Rees:87} due to its requirement of substantial primordial
vorticity \citep[although see][for an argument that some vorticity
is necessary]{Italians:04,Japanese:04}, the fields of interest
here are generated by the earliest stages of structure formation,
{\it requiring no new physics}.
Ions (taken to be protons, for simplicity) and electrons 
are treated as separate fluids, with opposite charges
but significantly different masses. The mass-weighted sums of 
their density and velocity fields will determine the evolution of the 
net baryonic component of the universe, and should agree with previous 
treatments \citep[e.g.][]{MaBert:95}.  The differences of their density 
and velocity fields, however, provides a description of a local charge
separation and a local current density, both of which contribute
to magnetic fields.  Since cosmological perturbations, which serve as 
seeds for structure formation, exist on all scales, 
it is expected that seed magnetic fields will be generated
on all scales by this mechanism.  As the magnetic fields generated
via this mechanism are in place at very early times, they possess
the advantage over other astrophysical mechanisms that these
fields will have optimally large amounts of time to be affected 
by amplification effects and various dynamos.  The remainder of this paper
focuses on calculating the magnitude of the magnetic fields
generated by this process and discussing their cosmological
ramifications.

\section{Cosmological Perturbations}


Although the early universe is isotropic and homogeneous to 
a few parts in $10^{5}$ \citep{BW:97}, it is these small density
inhomogeneities, predicted by inflation to occur on all scales
\citep{GP:85}, that lead to all of the structure observed in the
universe today.
As it is the early epoch of structure formation
that is of interest for the creation of magnetic fields, 
we calculate the evolution of inhomogeneities 
in the linear regime of structure formation.  The most
sophisticated treatment of cosmological perturbations in the
linear regime to date is that of \citet{MaBert:95}, which provides
evolution equations for an inhomogeneous universe containing 
a cosmological constant, dark matter, baryons, photons, and
neutrinos.  This section extends their treatment to encompass
separate proton and electron components.  The mass-weighted sum of
protons and electrons will recover the usual baryon component, 
whereas the difference of the density fields represents a charge 
separation, and the difference of the velocities is a net current.

The dynamics of any cosmological fluid can be obtained, in
general, from the linear Einstein equations \citep[see][for
earlier treatments]{PY:70,SW:80,WS:81}.  Although the choice of
gauge does not impact the results, the Conformal Newtonian gauge
leads to the most straightforward calculations.
When tensor modes are unimportant, the metric is
given by
\begin{equation}
\label{metric} ds^2 = a^2(\tau) [ -(1+2\psi) d\tau^2 + (1-2\phi)
dx^i dx_i]\mathrm{,}
\end{equation}
where $\psi \simeq \phi$ when gravitational fields are weak.
The linearized Einstein equations are then 
\begin{eqnarray}
\label{einstein}
&& k^2 \phi + 3 \, \frac{\dot a}{a}(\dot \phi + \frac{\dot a}{a}\, \psi)
= 4 \pi G a^2 \delta T^0_{\, \, 0} \mathrm{,}
\nonumber\\
&& k^2 \, (\dot \phi + \frac{\dot a}{a}\, \psi) 
= 4 \pi G a^2 (\bar \rho +\bar P) \theta \mathrm{,}
\nonumber\\
&& \ddot \phi + \frac{\dot a}{a}(\dot \psi + 2 \dot \phi) 
+ (\frac{2 \ddot a}{a} - \frac{\dot a^2}{a^2}) \psi 
+ {k^2 \over 3} (\dot \phi - \dot \psi) 
= {4\pi\over 3} G a^2 \delta T^i_{\;i} \mathrm{,}
\nonumber\\
&& k^2(\phi - \psi) = 12 \pi G a^2 (\bar \rho + \bar P) \sigma
\mathrm{,}
\end{eqnarray}
where 
$\sigma$ is the shear stress, which is negligible for
non-relativistic matter but important for photons and neutrinos.
A cosmological fluid that is either uncoupled to the other fluids 
or mass-averaged among uncoupled fluids in the early universe obeys 
\begin{eqnarray}
\label{evolution}
\dot \delta &=& -(1+w)(\theta - 3\dot \phi) - 3 \frac{\dot a}{a}
(c_s^2 - w) \delta \mathrm{,}
\nonumber\\
\dot \theta &=& -\frac{\dot a}{a}(1-3w)\theta - \frac{\dot
w}{1+w}\theta + \frac{c_s^2}{1+w}k^2 \delta
\nonumber\\
&&-k^2 \sigma + k^2 \psi \mathrm{,}
\end{eqnarray}
where $\delta$ is defined as the local density relative to its 
spatial average $(\delta \equiv {\delta \rho} / {\bar{\rho}})$; 
$\theta \equiv i k^j v_j$, where $v_j$ is the local peculiar
velocity; and $c_s$ is the sound speed of the fluid.

For individual components with additional inter-component interactions,
equation (\ref{evolution}) must be augmented to include these
interactions.  Examples of such interactions include momentum 
transfer between photons and charged particles and Coulomb 
interactions between protons and electrons.  For protons, electrons, 
and cold dark matter (CDM), an equation of state $w=0$ is assumed, 
and for radiation and neutrinos, $w=\frac{1}{3}$.
The master equations for each component of interest are computed
explicitly in subsections 3.1--3.5.
The evolution equations which appear in this section for baryons,
photons, light neutrinos, and cold dark matter are derived in much
greater detail, including photon polarization, hydrogen and helium 
recombination, in both synchronous and conformal Newtonian gauges, 
in \citet{MaBert:95}.

\subsection{Cold Dark Matter}

The cold dark matter component (denoted by the subscript $c$)
is collisionless and pressureless; its evolution can be simply read off 
from equation (\ref{evolution}) with $ w = c_s^2 = \sigma = 0 $ to be 
\begin{eqnarray}
\label{CDM}
\dot \delta_c = &-&\theta_c + 3\dot \phi \mathrm{,}
\nonumber\\
\dot \theta_c = &-&\frac{\dot a}{a}\theta_c + k^2 \psi \mathrm{.}
\end{eqnarray}
Any cold (i.e. nonrelativistic), collisionless component will behave 
according to the dynamics given by equation (\ref{CDM}).

\subsection{Light Neutrinos}

For massless (or nearly massless) particles, pressure is non-negligible, 
and the appropriate equation of state has $ w = \frac{1}{3}$.
Additionally, the shear term ($\sigma$) may be important as well.
The only accurate way to compute the evolution of such a component of 
the universe is by 
integration of the Boltzmann Equation, which is given for light 
neutrinos (denoted by subscript $\nu$) by
\begin{equation}
\label{Boltz}
\frac{\partial \mathcal{F}_\nu}{\partial \tau} + i k (\hat k \cdot \hat n)
\mathcal{F}_\nu = 4 [\dot \phi - i k (\hat k \cdot \hat n) \psi] \mathrm{,}
\end{equation}
in Fourier space.

The approximation that neutrinos are massless and uncoupled is
very good from an age of the universe of approximately $t \simeq 1$ s 
until the epoch of recombination.  The evolution equations for light
neutrinos are then
\begin{eqnarray}
\label{nus}
\nonumber\\
\dot \delta_\nu &=& -\frac43 \theta_\nu + 4 \dot \phi \mathrm{,}
\nonumber\\
\dot \theta_\nu &=& k^2 ( \frac14 \delta_\nu - \sigma_\nu + \psi ) \mathrm{,}
\nonumber\\
\dot \mathcal{F}_{\nu l} &=& \frac{k}{2l + 1} [l \mathcal{F}_{\nu (l-1)} -
(l+1)\mathcal{F}_{\nu (l+1)}] \mathrm{,}
\end{eqnarray}
where $\sigma_\nu$ is related to $\mathcal{F}_\nu$ by $2
\sigma_\nu = \mathcal{F}_{\nu 2}$, and the index $l$ governs the
final equation for $l \geq 2$.  $\mathcal{F}_{\nu l}$ is defined
by the expansion of the perturbations in the distribution
function, $\mathcal{F}_\nu$,
\begin{equation}
\label{Fnu}
\mathcal{F}_\nu \equiv \sum^\infty_{l=0} (-i)^l (2l+1) \mathcal{F}_{\nu l}
(k, \tau) P_l (\hat k \cdot \hat n) \mathrm{,}
\end{equation}
where $P_l (\hat k \cdot \hat n)$ are the Legendre polynomials.
Equations (\ref{Boltz}--\ref{Fnu}) are valid for any non-collisional 
species behaving as radiation.

\subsection{Photons}

Photons (denoted by subscript $\gamma$), although similar to light neutrinos,
evolve differently due to their large coupling to charged particles.
The differential scattering cross-section of photons with electrons, 
is given by 
\begin{equation}
\label{thomson}
\frac{d\sigma}{d\Omega} = \frac{3 \sigma_T }{16 \pi} 
(1 + \cos^2 \theta) \mathrm{,}
\end{equation}
where $\sigma_T$ is the Thomson cross-section \citep{JDJ:98}.
Photons also scatter with protons, but with a cross-section suppressed 
by a factor of the squared ratio of electron to proton mass, 
$ m_e^2 / m_p^2 \approx 3.0 \times 10^{-7} $.

The polarization-summed phase-space distribution for photons, 
$\mathcal{F}_\gamma$, is the same as the distribution function
for neutrinos (see equation \ref{Fnu}), with a non-zero difference 
between the two linear polarization components 
denoted by $\mathcal{G}_\gamma$.
The linearized collision operators for Thomson scattering
\citep{BE:84,BE:87,Kos:94,MaBert:95} yield the set of master
equations for photons,
\begin{eqnarray}
\label{photons}
\dot \delta_\gamma &=& - \frac43 \theta_\gamma + 4 \dot \phi \mathrm{,}
\nonumber\\
\dot \theta_\gamma &=& k^2 ( \frac14 \delta_\gamma - \sigma_\gamma )
 + k^2 \psi + a n_e \sigma_T (\theta_b - \theta_\gamma) \mathrm{,}
\nonumber\\
\dot \mathcal{F}_{\gamma 2} &=& \frac{8}{15} \theta_\gamma - \frac35 k
\mathcal{F}_{\gamma 3} - \frac95 a n_e \sigma_T \sigma_\gamma
\nonumber\\
& & + \frac{1}{10} a n_e \sigma_T (\mathcal{G}_{\gamma 0} +
\mathcal{G}_{\gamma 2}) \mathrm{,}
\nonumber\\
\dot \mathcal{F}_{\gamma l} &=& \frac{k}{2 l + 1} [l \mathcal{F}_{\gamma(l-1)}
- (l+1) \mathcal{F}_{\gamma (l+1)}] - a n_e \sigma_T \mathcal{F}_{\gamma l}
\mathrm{,}
\nonumber\\
\dot \mathcal{G}_{\gamma m} &=& \frac{k}{2m+1} [m \mathcal{G}_{\gamma (m-1)}
- (m+1) \mathcal{G}_{\gamma (m+1)}]
\nonumber\\
& & + a n_e \sigma_T \left[ \frac{1}{10}
\mathcal{F}_{\gamma m} - \frac25 \mathcal{G}_{\gamma m} \right] \mathrm{,}
\end{eqnarray}
where $\mathcal{F}_{\gamma 0} = \delta_\gamma$, $\mathcal{F}_{\gamma 1} =
4 \theta_\gamma / 3 k$, $\mathcal{F}_{\gamma 2} = 2 \sigma_\gamma$,
the indices $l$ and $m$ are valid for $l \geq 3$ and $m \geq 0$, and
the subscript $b$ denotes the baryonic component, which is the
mass-weighted sum of the electrons and protons.  Electron-photon
scattering is so dominant over proton-photon scattering as to render
the latter negligible, but the electron-proton coupling (via
electromagnetism) is sufficiently strong that, to leading order,
those two fluids move in kinetic equilibrium.

\subsection{Baryons}

The net behavior of the baryonic component can be derived from combining 
the mass-weighted contributions of the proton and electron fluids.
Both protons and electrons contain all of the terms present 
in the CDM equations (cf. Section 3.1), but additionally 
contain important terms arising from the Coulomb interaction 
and from Thomson scattering.
The coupling of the Coulomb interaction to density inhomogeneities 
can be calculated through a combination of the electromagnetic
Poisson equation,
\begin{equation}
\label{poisson} \del^2 \Phi = -\del \cdot \vec E = -4 \pi
\rho_C \mathrm{,}
\end{equation}
where $\rho_C$ is the electric charge density, and the Euler equation,
\begin{equation}
\label{euler}
\frac1a \frac{\partial (a \v)}{\partial t} 
 + \frac1a (\v \cdot \del) \v = - \frac1a \del\phi 
 - \frac{q}{m} \frac1a \del \Phi + \mathcal{C} \mathrm{,}
\end{equation}
with $q/m$ as the charge-to-mass ratio of the particle in question
and $\mathcal{C}$ the collision operator.
The Coulomb contribution appears as $4 \pi e a^2 (n_p - n_e) q_i / m_i$ 
in the evolution equation for $\dot \theta_i$, where $i$ denotes a 
species of particle with a mass $m_i$ and charge $q_i$.
The evolution equations are therefore 
\begin{eqnarray}
\label{electrons}
\dot \delta_e &=& - \theta_e + 3 \dot \phi \mathrm{,}
\nonumber\\
\noalign{\smallskip} 
\dot \theta_e &=& - \frac{\dot a}{a} \theta_e + c_s^2 k^2 \delta_e + k^2 \psi 
\nonumber\\
&&{}+ \Gamma_e (\theta_\gamma - \theta_e)
- \frac{4 \pi e^2 a^2}{m_e} (n_p - n_e) 
\mathrm{,}
\end{eqnarray}
for electrons (denoted by subscript $e$), and
\begin{eqnarray}
\label{protons}
\dot \delta_p &=& - \theta_p + 3 \dot \phi \mathrm{,}
\nonumber\\
\noalign{\smallskip} 
\dot \theta_p &=& - \frac{\dot a}{a} \theta_p + c_s^2 k^2 \delta_p + k^2 \psi
\nonumber\\
&&{} + \Gamma_p (\theta_\gamma - \theta_p)
+ \frac{4 \pi e^2 a^2}{m_p} (n_p - n_e) 
\mathrm{,}
\end{eqnarray}
for protons, denoted by subscript $p$, where $\Gamma$ characterizes the 
rate of momentum transfer due to photon scattering with charged particles.
The damping coefficient for electrons $(\Gamma_e)$ is given by 
\begin{eqnarray}
\label{Gamma_e}
\Gamma_e &\equiv& \frac{4 \rhobar_\gamma n_e \sigma_T a}{3 \rhobar_e} 
\,\mathrm{,}
\end{eqnarray}
and the analogous quantity for protons is smaller by a factor 
$ \Gamma_p/\Gamma_e = (m_e/m_p)^3 \simeq 1.6 \times 10^{-10} $.
Note the difference in the sign of the final terms in the equations
for $\dot \theta_e$ and $\dot \theta_p$, which will prove important below.

From equations (\ref{electrons}) and (\ref{protons}) for electrons 
and protons the dominant gravitational and electromagnetic combinations
can be constructed separately.  The remainder of this subsection details
the evolution of baryons in the linear regime of a perturbed universe. 
Baryonic matter can be treated as the combination of electrons and 
protons, thus the mass weighted sum of proton and electron overdensities 
gives rise to the baryonic perturbations, 
\begin{equation}
\label{barydef}
\delta_b \equiv \frac{m_e}{m_b}\delta_e + \frac{m_p}{m_b}\delta_p 
\mathrm{,} \qquad \theta_b \equiv \frac{m_e}{m_b}\theta_e + 
\frac{m_p}{m_b}\theta_p \mathrm{,}
\end{equation}
where $ m_b = m_p + m_e $.
The evolution of baryonic matter follows from a mass-weighted 
combination of the equations for electrons (\ref{electrons}) 
and protons (\ref{protons}).
To the extent that electrons and protons move together (the tight-coupling 
approximation) and the difference $ n_p - n_e $ is negligibly
small (for baryonic evolution 
this is an excellent approximation; see Section 3.5), 
the evolution equations for baryonic matter are 
\begin{eqnarray}
\label{baryons}
\dot \delta_b &=& - \theta_b + 3 \dot \phi \mathrm{,}
\nonumber\\
\dot \theta_b &=& -\frac{\dot a}{a} \theta_b + c_s^2 k^2 \delta_b 
+ k^2 \psi  + \Gamma_b (\theta_\gamma - \theta_b) 
\mathrm{,}
\end{eqnarray}
where $ \Gamma_b = (m_p \Gamma_p + m_e \Gamma_e)/m_b 
\approx \Gamma_e m_e / m_b $.
In the tight coupling approximation, the baryon-photon coupling term 
in equation (\ref{baryons}) is driven  by the electron-photon interaction, 
as has been shown by \citet{Harrison:70} and subsequent authors.
The equations in (\ref{baryons}) are identical to the equations for 
the evolution of baryonic inhomogeneities derived in \citet{MaBert:95}.

\subsection{Charge Separation}

From equations (\ref{electrons}) and (\ref{protons}), a charge 
{\it difference} component as well as a sum component can be obtained.
As the limits on a net electric charge asymmetry in the universe
are very strict \citep{OY:85,MR:02,CF:05}, 
we expect that any differences in densities and/or velocities of protons 
and electrons will not be strong enough to significantly alter the 
evolution of the other components of the universe.  It is therefore
expected that the analysis of 
baryonic matter, detailed in Section 3.4, will be unaffected by
differences in proton and electron densities and velocities.  However,
even a small separation may still have important 
cosmological implications.

The charge difference component (denoted by subscript $q$) is the
difference between the proton and electron components, such that
$\delta_q = \delta_p-\delta_e$ and $\theta_q = \theta_p - \theta_e$.
The gravitational potential does not affect the evolution of these 
quantities; gravity acts equivalently on electrons and protons.
However, if $ \delta_q $ and $ \theta_q$ are nonzero, 
resulting electromagnetic fields will act differently on 
oppositely charged particles.
The master equations for the charge-asymmetric component, obtained
from equations (\ref{electrons}) and (\ref{protons}) are 
\begin{eqnarray}
\label{A}
\dot \delta_q &=& - \theta_q	\nonumber\\
\dot \theta_q &=& - \frac{\dot a}{a} \theta_q + c_s^2 k^2 \delta_q
 + {4 \pi n_e e^2 a^2 \over m_e} \, \delta_q  \nonumber\\
&&{} - \Gamma_e (\theta_\gamma - \theta_b + \theta_q) 
\mathrm{,}
\end{eqnarray}
where the approximations $\Gamma_p \ll \Gamma_e$ and $m_b \simeq m_p$
have been utilized where applicable.
The term $ 4 \pi n_e e^2 a^2 \deltaq / m_e $ in equation (\ref{A}) 
arises from the Coulomb force acting on charged particles, 
while the final term, $\Gamma_e (\theta_\gamma - \theta_b + \theta_q)$, 
arises from the
difference in Thomson scattering between protons and electrons.
This final term provides a source for the generation of a charge 
separation independent of 
and in addition to any initial charge asymmetry, and will create a 
local charge asymmetry {\it even when there is none initially}.
In the evaluation of equation (\ref{A}), 
the electromagnetic terms dominate the cosmological terms, such
that an excellent approximation in the pre-recombination universe is
\begin{equation}
\label{AA}
\dot \delta_q = -\theta_q \mathrm{,} \qquad 
\dot \theta_q = {4 \pi n_e e^2 a^2 \over m_e} \, \delta_q
- \Gamma_e (\theta_\gamma - \theta_b + \theta_q) \mathrm{.}
\end{equation}

For some purposes, it is useful to express the set of equations found
in equation (\ref{A}) as a single ordinary differential equation.  
This can be accomplished by setting $\dot \theta_q = - \ddot \delta_q$, 
and again by neglecting the unimportant cosmological terms 
$ \theta_q \dot a / a $ and $ c_s^2 k^2 \delta_q $.  Many of the
coefficients in equation (\ref{A}) are functions of $a$, but the
derivatives in equation (\ref{A}) are with respect to conformal time,
$\tau$.  A change of variables can be performed, using the relation that
\begin{eqnarray}
t &=& \left[ \frac{45 \hbar^3 c^5}{32 \pi^3 G (kT)^4} \right]^{1/2}
\nonumber\\
&=& N a^2 t_0 \mathrm{,} \qquad  N \simeq 72.2 \mathrm{,}
\label{ttoa}
\end{eqnarray}
in the radiation era, where $t_0$ is the age of the universe today and
$t \equiv a \tau$,
to express all derivatives as derivatives with respect to $a$, denoted
by primes (instead of dots). 
The evolution of $\delta_q $ can be tracked by evolving equation 
(\ref{solve}) below,
\begin{eqnarray}
\label{solve}
\delta_q '' &+& 2 N \, \Gamma_{e,0} \frac{1}{a^2} \delta_q ' + 
\frac{16 N^2 \, \pi \, n_{e,0} \, e^2 a}{m_e} \delta_q 
\nonumber\\
&=& 4 N^2 \, \Gamma_{e,0} \frac{1}{a} ( \theta_\gamma - \theta_b ) \mathrm{,}
\end{eqnarray}
where the subscript $0$ denotes the present value of a quantity.
This is simply the equation of a damped harmonic oscillator, 
with coefficients that change slowly with time compared 
to damping or oscillation times.
The behavior can be characterized as overdamped at the earliest 
times, critically damped when $ a \approx 2.3 \times 10^{-9} $, 
and underdamped at late times.

Of all the terms in equation (\ref{solve}), only $\theta_\gamma$,
$\theta_b$, and $\delta_q$ (and derivatives) are functions of $a$; 
all other quantities are constant coefficients.  Although there does 
not exist a simple analytic form for $(\theta_\gamma - \theta_b)$ in
general, at sufficiently early times there exists the simple
approximation
\begin{equation}
\label{thetadiff} \theta_\gamma - \theta_b \simeq 6.0 \times
10^{19} \, k^4 \, a^5 \mathrm{,}
\end{equation}
valid when the following condition is met:
\begin{eqnarray}
\label{numberless}
a \, \lesssim \, 10^{-5} \quad
&\mathrm{for}& \quad k \, \leq \, 0.1 \, \Mpc^{-1}
\mathrm{,} \nonumber\\
a \, \lesssim \, 10^{-6} \left( \frac{1 \, \Mpc^{-1}}{k} \right) \quad 
&\mathrm{for}& \quad k \, \geq \, 0.1 \, \Mpc^{-1} \mathrm{.} \nonumber
\end{eqnarray}
Equation (\ref{thetadiff}) is an approximation for a flat $\Lambda$CDM
cosmology with cosmological parameters $ H_0 = 71 \km \s^{-1}\Mpc^{-1} $,
$\Omega_m = 0.27$, $\Omega_b = 0.044 $, and a Helium-4 mass fraction of
$Y = 0.248 $.  These parameters are used in all subsequent analyses 
for the calculation of cosmological quantities.

The approximation in equation (\ref{thetadiff}) breaks 
down at later times.
When this occurs, numerical methods must be used to obtain the quantity 
$(\theta_\gamma - \theta_b)$.
The software package \COSMICS{} \citep{Bert:95} is ideal for performing 
this computation, as it performs the numerical evolution of equations 
(\ref{CDM},\ref{nus},\ref{photons}, and \ref{baryons}) concurrently. 
Computational results for the quantities $\theta_\gamma $ and
$\theta_b$ are given by \COSMICS, which are valid at all times 
in the linear regime of structure formation.  It is found that 
when the approximation in equation (\ref{thetadiff}) breaks down, 
the quantity $(\theta_\gamma - \theta_b)$ grows 
more slowly initially, and proceeds to oscillate at a roughly 
constant amplitude at later times.  These oscillations in the
quantity $(\theta_\gamma - \theta_b)$ are closely related to 
the acoustic oscillations between baryons and photons observed 
in the cosmic microwave background \citep{wmap:03}.

Numerical integration of equation (\ref{solve}) can be 
accomplished in various ways \citep[see][for examples]{NRIC}.
At sufficiently late times (when $a \gg 2.3 \times 10^{-9}$), 
numerical results indicate that the quasi-equilibrium solution 
\begin{equation}
\label{dsoln}
\deltaq = \frac{ \sigma_T \, m_b }{ 3 \, \pi \, e^2} 
\left(\frac{ \bar{\rho}_{\gamma\mathrm{,}0} }
{ \bar{\rho}_{b,0} } \right) \, \left( \frac{1}{a^2} \right) 
\, (\theta_\gamma - \theta_b) \mathrm{,}
\end{equation}
obtained by neglecting the first two terms in equation (\ref{solve}), 
is an excellent approximation.
With $ \theta_\gamma $ and $ \theta_b $ given by \COSMICS{} in units 
of $ \Mpc^{-1} $, the prefactor in equation (\ref{dsoln}) can be
written as
\begin{equation}
\frac{ \sigma_T \, m_b }{ 3 \, \pi \, e^2 } 
\left(\frac{ \bar{\rho}_{\gamma\mathrm{,}0} }
{ \bar{\rho}_{b,0} } \right) 
= \frac89  \frac{m_b}{m_e}
\frac{ \Omega_{\gamma,0}}{\Omega_{b,0} } \, r_e 
\approx 1.67 \times 10^{-37} \, \Mpc, \label{constant}
\end{equation}
where $ r_e = e^2/m_e c^2 $ is the classical electron radius.
The quantity $\theta_q$ then follows directly from equation
(\ref{A}) to be
\begin{equation}
\label{tsoln}
\thetaq = - \frac{ \sigma_T \, m_b }{ 3 \, \pi \, e^2 } 
\left(\frac{ \bar{\rho}_{\gamma\mathrm{,}0} }
{ \bar{\rho}_{b,0} } \right) \, \left( \frac{1}{a^2} \right) \,
(\dot{\theta}_\gamma - \dot{\theta_b}) \mathrm{.}
\end{equation}
The solutions in equations (\ref{dsoln}) and (\ref{tsoln}) 
remain valid up to the epoch of recombination $(z \simeq 1089)$.
\includegraphics[width=\columnwidth]{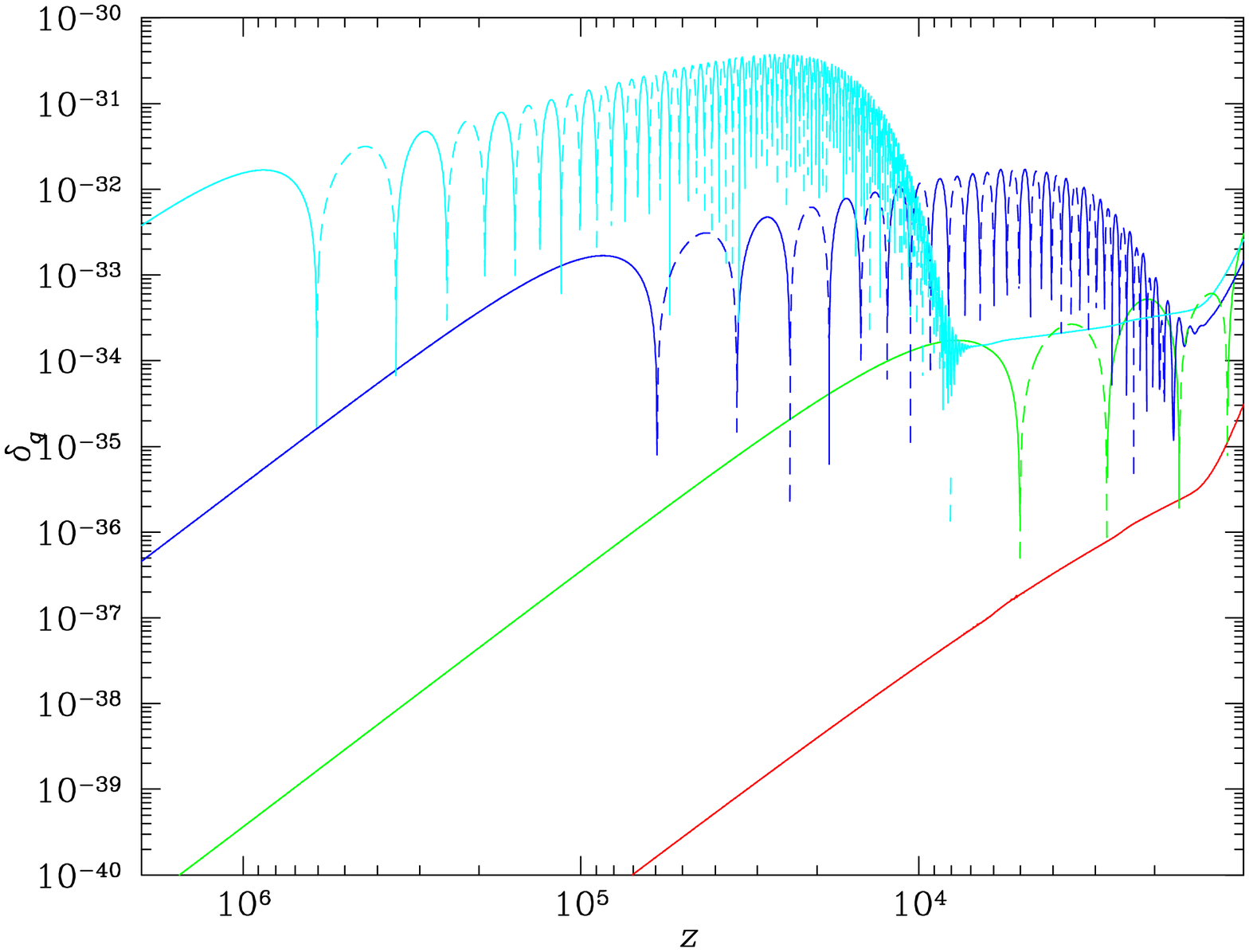}
\figcaption{Fourier mode amplitudes of the charge asymmetry $\deltaq$ 
plotted as a function of redshift $z$ for comoving scales of 
(from top to bottom) $ k = 10 $, 1, 0.1, and $ 0.01 \, \Mpc^{-1}$.
The amplitude $\deltaq$ rises as $\sim a^3$ initially, 
then ceases to grow when the scale of 
interest enters the horizon, and oscillates at an
amplitude which first continues to rise slowly, then falls, eventually
matching on to the equilibrium solution $\deltaq \propto \theta_b$.
Solid lines indicate positive values of $\delta_q$, 
and dashed lines indicate negative $\delta_q$. \label{DTfig}}

The results of numerically integrating the equations for 
$\delta_q$ and $\theta_q$ on various length scales through 
recombination  using the code \COSMICS{} \citep{Bert:95} 
are presented as a function of redshift $z$ in Figure \ref{DTfig} 
for $ k = 10 $, 1, 0.1, and $ 0.01 \, \Mpc^{-1}$.
The amplitude $\deltaq$ rises as $\sim a^5$ initially, 
then ceases to grow when the scale of 
interest enters the horizon, and oscillates at an
amplitude which first continues to rise slowly, then falls, eventually
matching on to the equilibrium solution $\deltaq \propto \theta_b$.  

The magnitude of the charge separation is small (of order 
$ 10^{-33} $ in Figure \ref{DTfig}) because the constant in 
equation (\ref{constant}) is small compared to the 
Hubble length at decoupling.
In the conformal Newtonian gauge the mode amplitudes are defined to be of 
order 1 outside the horizon, and need to be multiplied by the COBE 
normalization $ \delta_H = 1.95 \times 10^{-5} $ of \citet{BW:97} 
to obtain the physical quantities of magnetic field strength and
magnetic spectral density, as in Section 4.

Figure \ref{FigP_q} shows the spectral density of the charge asymmetry, 
$ 4 \pi k^3 P_q(k) / (2\pi)^3 $, plotted as a function of $k$.
For comparison, the spectral density of baryonic matter, 
$ 4 \pi k^3 P_b(k) / (2\pi)^3 $, is also shown.  Note the identical
scales, $k$, at which $P_q(k)$ and $P_b(k)$ change signs.

\includegraphics[width=\columnwidth]{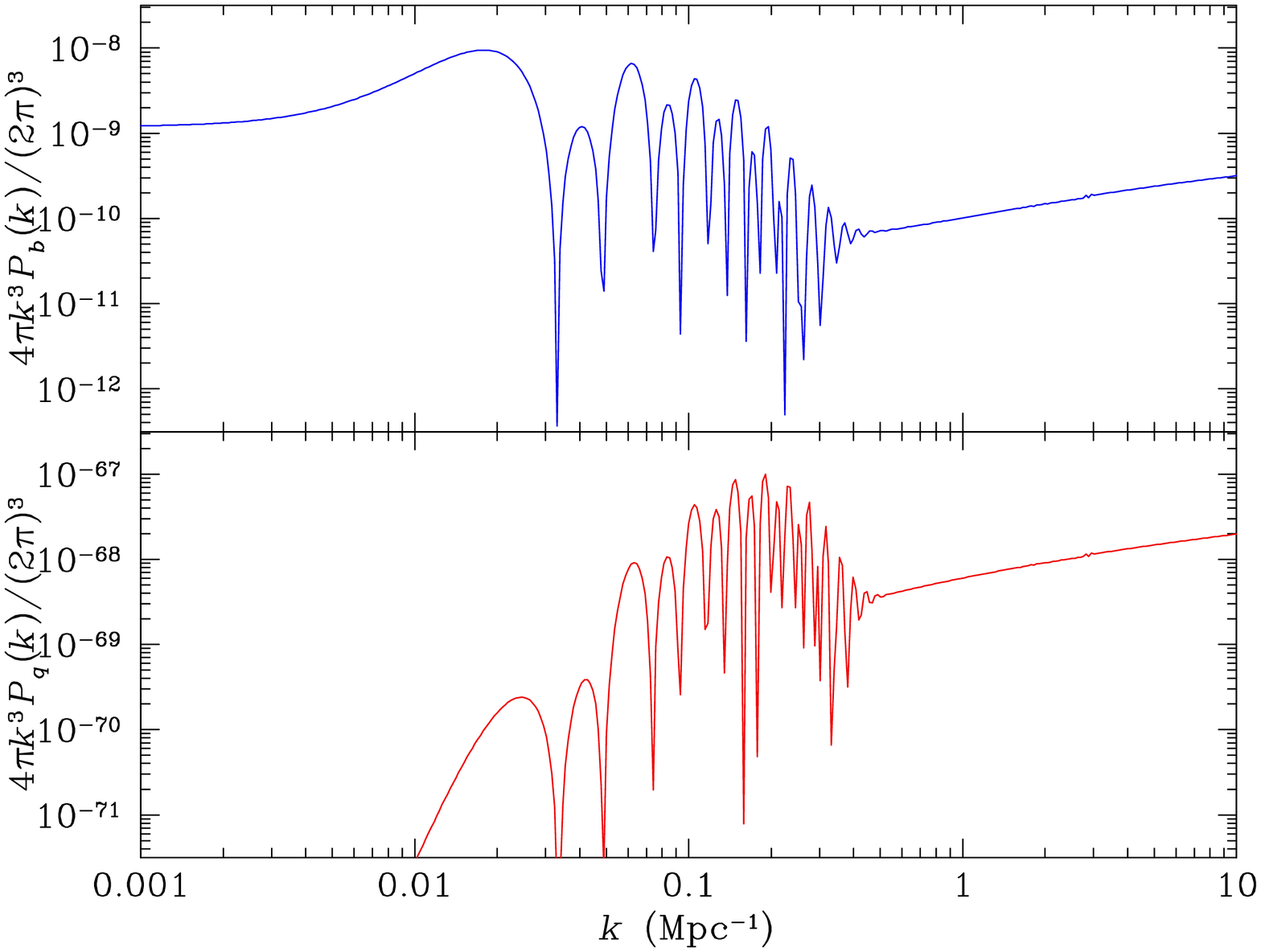}
\figcaption{Spectral density of the charge asymmetry, 
$ 4\pi k^3 P_q(k) /(2\pi)^3 $, 
at the epoch of decoupling, $ z = 1089 $, 
plotted as a function of wavenumber $k$ (lower graph).
Also shown, at the same epoch, is the  
spectral density of baryons, $ 4\pi k^3 P_b(k) /(2\pi)^3 $, 
(upper graph).
\label{FigP_q}}

The results thus far are accurate up through the epoch of recombination.
At this point, however, the universe transitions from a state with 
ionization fraction $\chi_e \simeq 1$ to a state where the ionization 
fraction is small, $\chi_e \approx 10^{-4}$ \citep{Peebles:68}, and
the universe is transparent to radiation.
The calculations of \citet{Italians:04} suggest that Silk damping 
affects only scales $ k > 1 \, \Mpc^{-1} $.
In the absence of interactions with photons, a 
charge separation would continue to evolve as 
\begin{equation}
\label{osc}
\ddot \delta_q = - {4 \pi \, n_{e} \, e^2 \, a^2 \over m_e} \, \delta_q 
= - {3 \over 2} \, \Omega_b \, H^2 \, K^2 \, a^2 \, \delta_q , 
\end{equation}
where $K^2$ is the ratio of the electric to gravitational forces,
$$
K^2 = {e^2 \over G m_p m_e} = 2.26 \times 10^{39} \mathrm{.} 
$$
A residual charge separation without additional interactions would oscillate
(plasma oscillations) with proper frequency $ \omega \approx KH $, 
with a further slow adiabatic decay of the oscillation amplitude.
However, there are many other effects that begin to become
important after recombination, including gravitational collapse,
dynamo effects, and continued electron-photon scattering, 
and we do not attempt to compute the subsequent evolution in detail
in this paper.


\section{Magnetic Fields}

With the results derived in Section 3.5 for $\thetaq$ and
$\deltaq$, values for the local current densities and local
charge separations can be obtained at any time in the 
pre-recombination universe on all scales. Both $\deltaq$ and $\thetaq$ 
will contribute to magnetic fields, as currents create
magnetic fields directly, and the bulk motion of a region of net
charge will also produce a magnetic field.  For each comoving
distance scale (given by the value of $k$) and each epoch 
(determined by the scale factor $a$) of the universe, there will
be a magnetic field amplitude associated with that scale.
This field may serve as the seed for the large-scale magnetic fields 
observed today.

An expression for magnetic fields can be derived from 
the currents arising from the relative motion of the protons and
electrons in the universe.
Magnetic fields can be derived from Maxwell's equations
\begin{eqnarray} \label{Maxwell}
\del \cdot \B = 0 , \qquad  
\del \times \B = {4 \pi} \vec J 
+ \frac{\partial \vec E}{\partial t} \mathrm{,}
\end{eqnarray}
with the current density $\vec J$ given by
\begin{equation} \label{J}
\vec J = n_p \, e \, \v_p - n_e \, e \, \v_e 
\simeq n_e e [\delta_q \v_b + (1+\delta_b) \v_q ]
\mathrm{,}
\end{equation}
where $\v_q \equiv \v_p - \v_e$, and the displacement current 
is neglected.

In the Fourier domain, the curl of Maxwell's equation for 
$ \del \times \B $ gives a direct expression for the magnetic field 
as a convolution 
\begin{eqnarray} \label{Bfield}
\B(\k) &=& \frac{4 \pi \, n_{e,0} \, e}{a^2 |\k|^2}
\int {d^3 k' \over (2\pi)^3} {\k \times \k' \over |\k'|^2} \nonumber\\
&& [\theta_b (\k') \, \delta_q (\k - \k') +
\theta_q (\k') \, \delta_b (\k - \k') ] \mathrm{,}  
\end{eqnarray}
from which the second moment of the magnetic field $\B(\k)$ is 
\begin{eqnarray}
\label{Bmoment}
\lexp B_i (\k_1) B_j (\k_2) \rexp &=& 
(2\pi)^3 \dD(\k_1 + \k_2) \, P_{ij}^{\perp} \, P_B (k) , 
\nonumber\\
P_{ij}^{\perp} &\equiv& \frac12 ( \delta_{ij} - \khat_i \khat_j )
\mathrm{,}
\end{eqnarray}
where $\dD$ is the Dirac delta function and $ P_B (k) $ is the 
magnetic field power spectrum.  Note that the direction parallel to 
$\k$ does not contribute to magnetic fields.  Therefore, the
direction perpendicular to $\k$ is projected out in equation 
(\ref{Bmoment}).  The power spectrum, $ P_B (k) $, is then given by
\begin{eqnarray} \label{P_B}
P_B(k) &=& \left( \frac{4 \pi \, n_{e,0} \, e}{a^2 |\k|^2} \right)^2 
\int {d^3 k' \over (2\pi)^3} \, |\k|^2 \, \sin ^2 \lambda' \nonumber \\
&&\times \Bigl[ { 1 \over |\k'|^2} \, P_{ \theta_q \theta_q }(|\k'|) 
 P_{ \delta_b \delta_b }(|\k - \k'| ) \nonumber\\
&& - {1 \over |\k-\k'|^2} \, P_{ \theta_q \delta_b }(|\k'|) 
P_{ \theta_q \delta_b }(|\k - \k'| ) \nonumber \\
&&+ { 1 \over |\k'|^2} \, P_{ \theta_b \theta_b }(|\k'|) 
 P_{ \delta_q \delta_q }(|\k - \k'| )  \nonumber\\
&&- {1 \over |\k-\k'|^2} \, P_{ \theta_b \delta_q }(|\k'|) 
 P_{ \theta_b \delta_q }(|\k - \k'| ) \nonumber \\
&&+ { 2 \over |\k'|^2} \, P_{ \theta_q \theta_b }(|\k'|) 
 P_{ \delta_q \delta_b }(|\k - \k'| ) \nonumber\\
&&- {2 \over |\k-\k'|^2} \, P_{ \theta_q \delta_q }(|\k'|) 
 P_{ \theta_b \delta_b }(|\k - \k'| ) \Bigr] \mathrm{,} 
\end{eqnarray}
where $\lambda'$ is the angle between the vectors $\k$ and  $\k '$. 
The second moment of two general quantities, $\phi$ and $\psi$, is expressed 
as the power $P_{\phi \psi} (k)$, defined by
\begin{equation}
\label{generalP}
\langle \tilde{\phi}(\k_1) \tilde{\psi}(\k_2) \rangle = (2 \pi)^3 
\dD (\k_1 + \k_2)\,  P_{\phi \psi} (k) \mathrm{,}
\end{equation}
where $ k = |\k_1| = |\k_2| $.
All terms in equation~(\ref{P_B}) make comparable contributions 
to the total.
Note that although the velocities $ \v_e $ and $ \v_p $ 
(or $ \v_b $ and $ \v_q $) are derived from potentials, 
they appear in convolution with density inhomogeneities.  For this reason,
there is no need for explicitly second-order quantities (such as vorticity
or anisotropic stress) to obtain a nonzero magnetic field, 
an effect similar to the calculation of the kinetic 
Sunyaev-Zel'dovich effect by \citet{OV:86}.

The spectral density, $ 4 \pi k^3 P_B(k) / (2 \pi)^3 $, obtained by 
numerically integrating $\theta_q$ and $\delta_q$ in equations 
(\ref{dsoln}) and (\ref{tsoln}),
provides both a measure of the magnetic field strength on a given 
scale $k$ and a measure of the energy stored in magnetic fields, 
$ \rho_B = |\vec{B}|^2/8\pi $.
The results for the spectral density of the magnetic field on 
comoving scales from $10^{-3} \,\Mpc^{-1}$ to $10^2 \,\Mpc^{-1}$ 
at the epoch of recombination are shown in Figure \ref{Bgraph}.
The peak of the density corresponds to a typical magnetic 
field strength of $10^{-25}$--$10^{-24} \, \G $ on a comoving scale 
of $0.1 \,\Mpc^{-1}$.

\includegraphics[width=\columnwidth]{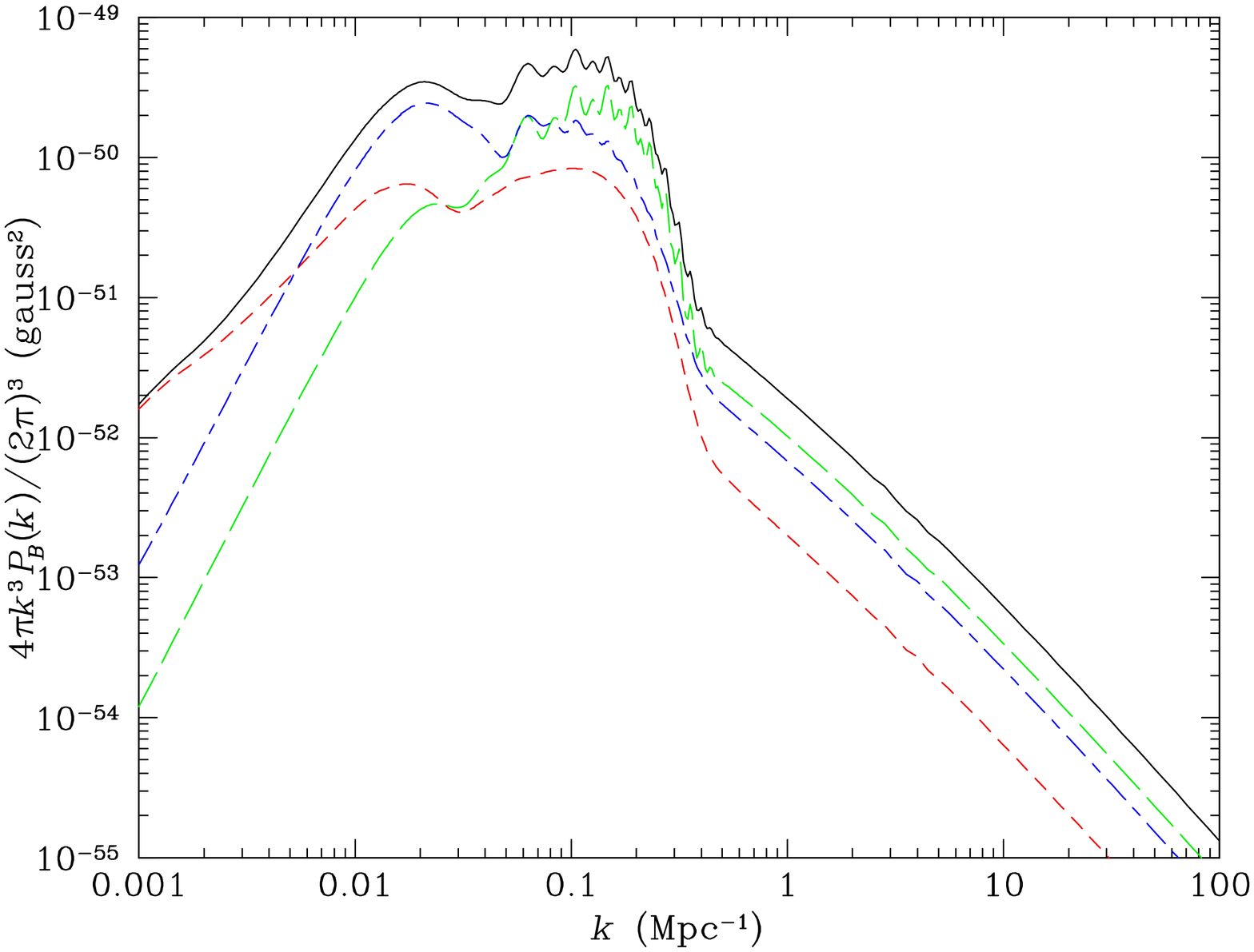}
\figcaption{Spectral density of the magnetic field 
$ 4 \pi k^3 P_B(k)/ (2\pi)^3 $ (in $\G^2$) generated 
by cosmological perturbations on a given comoving scale $k$ 
(in $ \Mpc^{-1}$) at the epoch of recombination $z \simeq 1089$.
The dashed lines show the three separate contributions 
in equation (\ref{P_B}) (where the term containing $ P_{\theta_q \theta_q} $ 
is short-dashed; $ P_{\theta_b \theta_b} $ is long-dashed; 
and $ P_{\theta_q \theta_b} $ is short/long-dashed), 
and the solid (topmost) line shows the total magnetic spectral density.
The peak value corresponds to a magnetic field strength 
$ B \sim 10^{-25} $--$ 10^{-24}\,\G $.
Note also that the oscillations due to Silk damping become 
unimportant for $k \gtrsim 0.5 \,\Mpc^{-1} $.
\label{Bgraph}}

\section{Discussion}


The major result of this paper has been to demonstrate that seed magnetic
fields of cosmologically interesting strengths and scales are
necessarily generated by the same processes that cause structure
formation.
As overdensities in the early universe slowly grow during the 
radiation era, the differing photon interactions with protons and 
electrons create charge separations and currents of small magnitudes 
on all scales.
These charge separations and currents grow in magnitude as 
the universe evolves, causing magnetic fields to grow as well.
The magnetic power in a given mode peaks at approximately the time of 
horizon crossing, oscillating and falling slowly in amplitude after that.
The net result is that, at the epoch of recombination 
(and hence prior to any field amplification due to 
gravitational collapse or dynamo effects), seed magnetic fields 
of magnitude  $\mathcal{O}(10^{-24} \, \G) $
are created by the simple dynamics of charged particles.

Other recent discussions of the origins of large-scale magnetic fields 
have mentioned the possibility that the evolution of cosmological
perturbations could be responsible for their generation 
\citep{Italians:04,Japanese:04,GS:05,Ichiki:06}.  These groups 
have focused on the 
effects of vorticity, which arises at second order in cosmological
perturbation theory.  
In addition to the charge separations and currents induced by 
photon scattering and the Coulomb interaction, as discussed in this
work, \citet{Japanese:04}
also considered the contribution of photon anisotropic stress 
to magnetic fields, but found that it is uninteresting on 
large scales.
The overall strength and scale of the magnetic fields reported by 
these authors seem to differ by large factors from one another. 
The fields derived in \citet{Japanese:04} and \citet{Ichiki:06} appear 
to be unreasonably large, and peak at scales where Silk damping ought to 
be a dominant effect \citep{Italians:04}.  As shown in the subsequent 
paragraph, their value for a magnetic field strength of $ B \sim 10^{-19} \,
\G $ on $\mathcal{O}( \Mpc )$ scales is difficult to reconcile with
other estimates.  It also appears that the fields derived in
\citet{GS:05} are insignificantly small, as their analysis of vorticity 
may not include the dominant terms in the creation of magnetic fields.
For reasons discussed in the next paragraph, it appears that the 
estimates for magnetic fields generated by vorticity are most
accurately given by \citet{Italians:04}.
When their results are considered 
on $\sim 10 \, \Mpc$ scales,
the fields generated are found to be of physically reasonable strengths 
similar to our own, of order $\sim 10^{-25} \, \G $.  Their spectrum is
similar to the spectrum of Figure \ref{Bgraph} as well, although 
their spectrum peaks at a slightly higher amplitude,
is narrower in shape, and reaches its maximum at a smaller scale.

The approximate magnitude of the field strengths arising from the 
full calculations of Section 4 can be understood intuitively.
From the quasi-equilibrium solution in equation~(\ref{dsoln}), the 
fractional charge asymmetry $(\delta_q)$ is of order $ 10^{-33} $ at 
the scale of the peak of the power spectrum. 
When this information is combined with the fact that the streaming velocity at 
decoupling is $ v/c \sim 10^{-5} $, a current that yields magnetic field
strengths of $\mathcal{O} ( 10^{-25} \, \G)$ is produced.
The field strengths expected from vorticity are somewhat less intuitive.
Although for an ideal fluid a rotational velocity is not generated from 
an initially longitudinal velocity if there is no vorticity at 
linear order \citep{Hwang:93}, 
this does not remain the case if the imperfect nature of the fluid 
is properly taken into account \citep{Vishniac:82}.
A reasonable ``upper-bound'' estimate of the magnetic field that 
arises from vorticity can be calculated from the angular momentum
of a protogalaxy, yielding an angular velocity $(\omega)$ 
and therefore a magnetic field.
\citet{Harrison:70} showed that magnetic fields of magnitude 
$ B \sim 10^{-4} \omega \, \G $ would arise in the radiation era due to 
vorticity.
From \citet{Peebles:69}, an estimate of the angular momentum in the 
baryonic component of galaxies is $\sim 10^{70} \g \cm^2 \s^{-1} $,
which yields a magnetic field of $ B \sim 10^{-22} \, \G $ 
for a protogalaxy at recombination.
This estimate of the field strength is reasonable as an 
upper bound, as the angular momentum from
\citet{Peebles:69} is for the entire protogalactic system, but the 
associated magnetic field energy may 
appear distributed across many smaller-scale modes.
Furthermore, this estimate is comparable to and consistent with 
the field strengths found
in \citet{Italians:04}.  It is therefore our assessment that 
their paper is the most correct of the works detailing 
magnetic fields arising from the vorticity generated by 
cosmological perturbations. 

It should not be surprising that the calculational frameworks of this
paper and those involving magnetic fields arising from vorticity can
lead to comparable results, as both calculations rely on second order
quantities.
The calculation of a vorticity in the context of cosmological perturbation
theory requires a velocity term that is explicitly second order, 
whereas Section 4 of this paper requires the convolution of 
two first order quantities.  On dimensional grounds, the two methods should 
yield comparable results.
The results in this paper rely only on fields known to exist, 
and whether the contribution from charge separations and currents or from
vorticity dominates can vary both with scale and from system to system.
The usual evolution of cosmological perturbations leads in a 
straightforward manner to charge separations and currents, 
which necessarily arise from the differing interactions between 
protons and electrons with photons. 
As has been shown in Section 4, magnetic fields follow directly 
from these sources.

A seed field of $ \mathcal{O}(10^{-24} \, \G )$, as generated
via the charge separations and currents induced by structure formation,
would provide an excellent candidate for the origins of cosmological
magnetic fields.  While field amplification due to gravitational collapse is 
negligible at the epoch of recombination, this will not be the case
at all times.  At recombination, the universe has only been
matter-dominated for a brief time, and thus density perturbations have
only grown by a small amount in that time, leading to an insignificant
amplification of the field strength.  As magnetic flux is frozen in, 
nonlinear collapse causes $|\vec B|$ to increase by many
orders of magnitude \citep{LC:95,HK:97}. 
However, the major source of amplification of an initial seed field 
comes from dynamo effects, as discussed in Section 2.  The key
to solving the puzzle of the origin of cosmic magnetic fields lies 
in determining whether the seed fields produced by a given mechanism
can be successfully amplified into the $\mathcal{O}(\mu \mathrm{G})$
fields observed today.  A major problem with many of the mechanisms 
that produce seed fields is that they produce weak fields at 
times insufficiently early for dynamo amplification to produce 
fields as large as $\sim \mu $G.  The Biermann mechanism, for
instance, can produce seed fields of order $\sim 10^{-19}$ G, but only
at a redshift of $z \sim 20$.  Although those initial fields are larger
than the $\sim 10^{-24}$ G fields produced by the growth of cosmic
structure, the fact that magnetic fields from structure formation 
are in place at $z \simeq 10^3 $ makes them an extremely attractive 
candidate for the seeds of cosmic magnetic fields.  As argued by 
\citet{DLT:99}, a seed field as small as $10^{-30}$ G at recombination
could possibly be amplified into a $\mu$G field today.  Clearly,
more work on understanding dynamo amplification is necessary before
a definitive solution to the puzzle of cosmic magnetic fields can
emerge.

One interesting mechanism worth investigating further is for the cosmic
seed fields generated by density perturbations to seed supermassive
black holes.  It is known that the magnetic field energy in active galactic 
nuclei and quasars is comparable to the magnetic field energy in 
an entire galaxy.  However, these structures cannot generate their own
magnetic fields from nothing; they require a pre-existing seed field.  
It therefore appears to be a reasonable 
possibility that the seed fields generated by cosmic structure
formation could provide the necessary fields to seed supermassive black
holes.  The resultant amplification via collapse and dynamo effects
could explain the origin of large-scale magnetic structures in the
universe. 

If large-scale magnetic fields exist at the epoch of recombination,
they may be detectable by upcoming experiments.  The results shown 
in Figure \ref{Bgraph} provide a prediction of
large-scale magnetic fields at the epoch of the cosmic microwave
background.  Sufficiently large magnetic fields on large scales at
recombination may be detectable by PLANCK \citep{Lewis:04,Planck:05}, 
although current estimates of their 
sensitivity indicate that the field strengths predicted in this paper
$( \sim 10^{-24} \, \G )$ would be significantly out of range 
of PLANCK's capabilities $( \sim 10^{-10} \, \G )$.  
Nonetheless, a knowledge of the field strengths at recombination allow
for predictions of CMB photon polarizations and Faraday rotation, both 
of which may be, at least in principle, observable.  

It is also of interest to note that any primordial charge asymmetry 
or large-scale currents (and therefore magnetic fields) created in
the very early universe will be driven away by these dynamics. 
Equation (\ref{solve}) has an approximate solution for $\deltaq$ 
which is critically (exponentially) damped at redshift 
$ z \simeq 10^{9}$, 
capable of reducing any pre-existing charge or current 
by as much as a factor of $\sim 10^{-10^{17}} $. 
As this factor is extraordinarily large, it is easy to conclude that
any initial $\delta_q$ or $\theta_q$ will be driven quickly to
the value given by equations (\ref{dsoln}) and (\ref{tsoln}) at the 
epoch of critical damping.
This result is independent of $k$ and ought to be applicable even 
to a charge asymmetry on large scales, perhaps approaching that 
of the present horizon.
As a result of this reduction of pre-existing charges or currents, 
the late-time $( z \lesssim 10^{9} )$ solutions obtained in this 
paper for charge separations, currents,
and magnetic fields ought to be independent of the initial conditions for
$\delta_q$, $\theta_q$, and $|\vec B|$ in the early universe.

Overall, the dynamics of ions, electrons, and photons during
the radiation era necessarily leads to charge separations and
currents on all scales, which in turn generate magnetic fields.
These fields supersede any pre-existing fields and are in place 
prior to substantial gravitational collapse.  Thus, the creation
of charge separations and currents from the evolution of cosmological 
perturbations emerges as a
promising and well-motivated new candidate to explain the 
origins of cosmic magnetic fields.

\acknowledgements

We thank Ed Bertschinger for providing the source code to the 
program \COSMICS, which proved invaluable in our analysis, and Steve
Detweiler for his input towards solving the differential equations
in this paper.  We also thank Ethan Vishniac for his extremely 
helpful comments on the first draft of this paper. 
E.R.S. acknowledges the University of Florida's
Alumni Fellowship program for partial support.
This research has made use of NASA's Astrophysics Data System.



\begin{thebibliography}{}

\bibitem[Ashoorioon \& Mann(2005)]{AM:05}
Ashoorioon, A., \& Mann, R.~B.\ 2005, \prd, 71, 103509

\bibitem[Athreya et al.(1998)]{Athreya:98} 
Athreya, R.~M., Kapahi, P.~J., McCarthy, P.~J., \& van Breugel, W.\ 1998, 
Astron. Astrophys.\ 329, 809

\bibitem[Barrow et al.(1997)]{BFS:97}
Barrow, J.~D., Ferreira, P.~G., \& Silk, J.\ 1997, \prl, 78, 3610

\bibitem[Baym et al.(1996)]{BBM:96}
Baym, G., B\"{o}deker, D., \& McLerran, L.\ 1996, \prd, 53, 662

\bibitem[Bennett et al.(2003)]{wmap:03}
Bennett, C.~L.\ et al.\ 2003, \apjs, 148, 1

\bibitem[Bertschinger (1995)]{Bert:95}
Bertschinger, E., arXiv:astro-ph/9506070.

\bibitem[Biermann(1950)]{Biermann:50}
Biermann, L.\ 1950, Z. Naturforsch., 5A, 65

\bibitem[Bond \& Efstathiou(1984)]{BE:84} Bond, J.~R., \&
Efstathiou, G.\ 1984, \apjl, 285, L45

\bibitem[Bond \& Efstathiou(1987)]{BE:87} Bond, J.~R., \&
Efstathiou, G.\ 1987, \mnras, 226, 655

\bibitem[Bunn \& White(1997)]{BW:97} Bunn, E.~F., \& White,
M.\ 1997, \apj, 480, 6

\bibitem[Calzetta et al.(1998)]{CKM:98} Calzetta, E.~A., Kandus,
A., \& Mazzitelli, F.~D.\ 1998, \prd, 57, 7139

\bibitem[Caprini \& Ferreira(2005)]{CF:05} Caprini, C., \&
Ferreira, P.~G.\ 2005, JCAP, 2, 6

\bibitem[Cheng et al.(1994)]{CST:94} Cheng, B., Schramm, D.~N., \&
Truran, J.~W.\ 1994, \prd, 49, 5006

\bibitem[Clarke et al.(2001)]{CKB:01} Clarke, T.~E., Kronberg,
P.~P., \& B\"{o}hringer, H.\ 2001, \apj, 547, 111

\bibitem[Davis et al.(2001)]{Davis:01} Davis, A.-C., Dimopoulos,
K., Prokopec, T., \& T\"{o}rnqvist, O.\ 2001, Phys. Lett. B, 501,
165

\bibitem[Davis et al.(1999)]{DLT:99} Davis, A.-C., Lilley, M., \&
T\"{o}rnkvist, O., 1999, \prd, 60, 021301

\bibitem[Delabrouille(2004)]{Planck:05} Delabrouille, J.\ 2004,
\apss, 290, 87

\bibitem[Fitt \& Alexander(1993)]{Fitt:93} Fitt, A.~J., \&
Alexander, P.\ 1993, \mnras, 261, 445

\bibitem[Gnedin et al.(2000)]{Gnedin:00} Gnedin, N.~Y., Ferrara,
A., \& Zweibel, E.~G.\ 2000, \apj, 539, 505

\bibitem[Gopal \& Sethi(2005)]{GS:05}
Gopal, R., \& Sethi, S.~K.\ 2005, \mnras, 363, 521 

\bibitem[Guth \& Pi(1982)]{GP:85} Guth, A.~G., \& Pi, S.-Y.\ 1982,
\prl, 49, 1110

\bibitem[Harrison(1970)]{Harrison:70} Harrison, E.~R.\ 1970,
\mnras, 147, 279

\bibitem[Harrison(1973)]{Harrison:73} Harrison, E.~R.\ 1973,
\prl, 30, 18

\bibitem[Hogan(1983)]{Hogan:83} Hogan, C.~J.\ 1983, \prl, 51, 1488

\bibitem[Howard \& Kulsrud(1997)]{HK:97} Howard, A.~M., \& Kulsrud,
R.~M.\ 1997, \apj, 483, 648

\bibitem[Hoyle(1969)]{Hoyle:69} Hoyle, F.\ 1969, Nature (London),
223, 936

\bibitem[Hwang(1993)]{Hwang:93}
Hwang, J.-C.\ 1993, \prd, 48, 3557 

\bibitem[Ichiki et al.(2006)]{Ichiki:06}
Ichiki, K., Takahashi, K.,Ohno, H., Hanayama, H., \& Sugiyama, N. 2006, 
Science, 311, 827



\bibitem[Jackson(1998)]{JDJ:98}
Jackson, J.~D.\ 1998, Classical Electrodynamics, 3rd Edition, 
(New York: Wiley) 

\bibitem[Kandus et al.(2000)]{Kandus:00} Kandus, A., Calzetta,
E.~A., Mazzitelli, F.~D., \& Wagner, C.~E.~M.\ 2000, Phys. Lett.
B, 472, 287

\bibitem[Kazantsev et al.(1985)]{Kazantsev:85} Kazantsev, A.~P.,
Ruzmaikin, A.~A., \& Sokoloff, D.~D.\ 1985, Zh. Eksp. Teor. Fiz.\
88, 487

\bibitem[Kim et al.(1990)]{Kim:90} Kim, K.-T., Kronberg, P.~P.,
Dewdney, P.~E., \& Landecker, T.~L.\ 1990, \apj, 355, 29

\bibitem[Kim et al.(1989)]{Kim:89} Kim, K.-T., Kronberg, P.~P.,
Giovannini, G., \& Venturi, T.\ 1989, Nature (London) 341, 720

\bibitem[Klein et al.(1988)]{Klein:88} Klein, U., Wielebinski, R.,
\& Morsi, H.~W.\ 1988, Astron. Astrophys.\ 190, 41

\bibitem[Kosowsky(1996)]{Kos:94}
Kosowsky, A.\ 1996, Annals Phys.\  246, 49

\bibitem[Kronberg et al.(1992)]{KPZ:92}
Kronberg, P.~P., Perry, J.~J., \& Zukowski, E.~L.~H.\ 1992, \apj, 387, 528

\bibitem[Kulsrud et al.(1997)]{Kulsrud:97} Kulsrud, R.~M., Cen,
R., Ostriker, J.~P., \& Ryu, D.\ 1997, \apj, 480, 481

\bibitem[Lesch \& Chiba(1995)]{LC:95} Lesch, H., \& Chiba, M.\
1995, Astron. Astrophys.\ 297, 305

\bibitem[Lewis(2004)]{Lewis:04} Lewis, A.\ 2004, \prd, 70,
043011

\bibitem[Ma \& Bertschinger(1995)]{MaBert:95} Ma, C.-P., \&
Bertschinger, E.\ 1995, \apj, 455, 7

\bibitem[Mass{\'o} \& Rota(2002)]{MR:02} Mass{\'o}, E., \& 
Rota, F.\ 2002, Phys. Lett. B, 545, 221 

\bibitem[Matarrese et al.(2005)]{Italians:04} Matarrese, S.,
Mollerach, S., Notari, A., \& Riotto, A.\ 2005, \prd, 71, 043502


\bibitem[Moffatt(1978)]{Moffatt:78} Moffat, H.~K.\ 1978,
Magnetic Field Generation in Electrically Conducting Fluids 
(Cambridge, Cambridge University Press)

\bibitem[Moss \& Shukurov(1996)]{Moss:96} Moss, D., \& Shukurov,
A.\ 1996, \mnras, 279, 229

\bibitem[Orito \& Yoshimura(1985)]{OY:85} Orito, S., \& 
Yoshimura, M.\ 1985, \prl, 54, 2457 

\bibitem[Ostriker \& Vishniac(1986)]{OV:86}
Ostriker, J. P., \& Vishniac, E. T. 1986, \apjl, 306, L5

\bibitem[Parker(1971)]{Park:71}
Parker, E.~N.\ 1971, \apj, 163, 255

\bibitem[Peebles(1968)]{Peebles:68} Peebles, P.~J.~E.\ 1968, \apj, 
153, 1 

\bibitem[Peebles(1969)]{Peebles:69} Peebles, P.~J.~E.\ 1969, \apj, 
155, 393 

\bibitem[Peebles \& Yu(1970)]{PY:70} Peebles, P.~J.~E., \&
Yu, J.~T.\ 1970, \apj, 162, 815

\bibitem[Press et al.(1992)]{NRIC} 
Press, W.~H., Teukolsky, S.~A., Vetterling, W.~T., \& Flannery, B.~P.\ 1992, 
Numerical Recipes in C (2nd ed.; Cambridge: Cambridge University Press)

\bibitem[Pudritz \& Silk(1989)]{Pudritz:89} Pudritz, R.~E., \&
Silk, J.\ 1989, \apj, 342, 650

\bibitem[Quashnock et al.(1989)]{QLS:89} Quashnock, J., Loeb, A.,
\& Spergel, D.\ 1989, \apjl, 344, L49

\bibitem[Ratra(1992)]{Ratra:92} Ratra, B.\ 1992, \apjl, 391, L1

\bibitem[Rees(1987)]{Rees:87} Rees, M.~J.\ 1987, \qjras, 28, 197

\bibitem[Ruzmaikin et al.(1988)]{Ruz:88} Ruzmaikin, A.~A.,
Shukurov, A.~M., \& Sokoloff, D.~D.\ 1988, Moscow, Izdatel'stvo
Nauka, 1988, 280 p.~In Russian.



\bibitem[Sigl et al.(1997)]{SOJ:97} Sigl, G., Olinto, A.~V., \&
Jedamzik, K.\ 1997, \prd, 55, 4582

\bibitem[Silk \& Wilson(1980)]{SW:80} Silk, J., \& Wilson,
M.~L.\ 1980, Phys. Scripta, 21, 708

\bibitem[Steenbeck et al.(1966)]{Steenbeck:66} Steenbeck, M.,
Krause, F., \& R\"{a}dler, K.~H.\ 1966, Z. Naturforsch. A, 21, 369

\bibitem[Subramanian et al.(1994)]{SNC:94} Subramanian, K.,
Narasimha, D., \& Chitre, S.~M.\ 1994, \mnras, 271, L15

\bibitem[Syrovatskii(1970)]{Syro:70} Syrovatskii, S.~I., in
Habing, H.~J.\ 1970, IAU Symp.~39: Interstellar Gas Dynamics, 39,
p. 192

\bibitem[Takahashi et al.(2005)]{Japanese:04}
Takahashi, K., Ichiki, K., Ohno, H., \& Hanayama, H.\ 2005, 
\prl, 95, 121301

\bibitem[Taylor et al.(1994)]{TBG:94} Taylor, G.~B., Barton,
E.~J., \& Ge, J.\ 1994, Astron. J.\ 107, 1942

\bibitem[Tribble(1993)]{Tribble:93} Tribble, P.~C.\ 1993, \mnras,
263, 31

\bibitem[Turner \& Widrow(1988)]{Turner:88} Turner, M.~S., \&
Widrow, L.~M.\ 1988, \prd, 37, 2743

\bibitem[Vainshtein \& Ruzmaikin(1971)]{Vain:71} Vainshtein,
S.~I., \& Ruzmaikin, A.~A.\ 1971, \azh, 48, 902

\bibitem[Vainshtein \& Ruzmaikin(1972)]{Vain:72} Vainshtein,
S.~I., \& Ruzmaikin, A.~A.\ 1972, \azh, 49, 449

\bibitem[Vall\'ee(1997)]{Vallee:97} Vall\'ee, J.~P.\ 1997, Fundam.
Cosmic Phys.\ 19, 1

\bibitem[Vishniac(1982)]{Vishniac:82} Vishniac, E.~T.\ 1982, \apj, 
253, 457 

\bibitem[Widrow(2002)]{Widrow:02} Widrow, L.~M.\ 2002, Rev.\ Mod.\
Phys.\ 74, 775

\bibitem[Wilson \& Silk(1981)]{WS:81} Wilson, M.~L., \& Silk,
J.\ 1981, \apj, 243, 14

\bibitem[Yamazaki et al.(2005)]{Yamazaki:05} Yamazaki, D.~G., 
Ichiki, K., \& Kajino, T.\ 2005, \apjl, 625, L1 

\bibitem[Yamazaki et al.(2006)]{Yamazaki:06} Yamazaki, D., Ichiki, 
K., Kajino, T., \& Mathews, G.~J.\ 2006, ArXiv Astrophysics e-prints, 
arXiv:astro-ph/0602224 

\bibitem[Zel'dovich \& Novikov(1983)]{ZN:83} Zel'dovich, Ya.~B.,
\& Novikov, I.~D.\ 1983, Chicago, IL, University of Chicago Press,
1983, 751 p.~Translation.

\end{thebibliography}
\end{document}